\documentclass[structabstract]{aa}  
\usepackage{graphicx}
\usepackage{txfonts}
\begin{document}
\title{Ion chemistry in the early universe:}

\subtitle{revisiting the role of HeH$^+$ with new quantum calculations}

\author{Stefano Bovino \inst{1}, Mario Tacconi \inst{1} Francesco A.
Gianturco\inst{1}\fnmsep\thanks{Corresponding author email:
fa.gianturco@caspur.it} \and Daniele Galli \inst{2} }

\institute{Department of Chemistry, ``Sapienza'' University of Rome, P.le A. Moro 5, 00185 Rome\\ 
\email{s.bovino@caspur.it} \and INAF-Osservatorio Astrofisico di Arcetri,\\
Largo E. Fermi 5, 50125 Firenze, Italy}

\date{Received ; accepted}

\abstract
%
{}
{The role of HeH$^+$ has been newly assessed with the aid of newly
calculated rates which use entirely ab initio methods, thereby allowing
us to compute more accurately the relevant abundances within the global
chemical network of the early universe.  A comparison with the similar
role of the ionic molecule LiH$^+$ is also presented.}
%
{Quantum calculations have been carried out for the gas-phase reaction
of HeH$^+$ with H atoms with our new in-house code, based on the
negative imaginary potential method. Integral cross sections and
reactive rate coefficients obtained under the general conditions of
early universe chemistry are presented and discussed.}
%
{With the new reaction rate, the abundance of HeH$^+$ in the early
universe is more than one order of magnitude larger than in previous
studies. Our more accurate findings further buttress the possibility 
to detect cosmological signatures of HeH$^+$.}
%
{}


\maketitle

\section{Introduction}

The existence of HeH$^+$  in astrophysical environments has been
extensively discussed in several papers.  Roberge \& Dalgarno~(1982)
surmised that HeH$^+$ is produced in planetary nebulae and in dense
molecular clouds under the presence of X-Ray and extreme UV ionization
sources. Liu et al.~(1997) investigated the possible detection of
HeH$^+$ in NGC~7027, a very high electron density nebular object, via
its $j$ = 1--0 strong rotational line at 149.14 $\mu$m, concluding that
its identification is complicated by the accidental near-coincidence of
that transition frequency with the 149.09 and 149.39 $\mu$m lines of
CH.

In the last years the presence of HeH$^+$  in low temperature and low
density helium-rich white dwarfs has further been investigated:  Harris
et al. (2004) demonstrated that this molecule is the dominant positive
ion in such objects and strongly affects their opacity.  Engel et
al.~(2005), finally, showed that it could also be possible to detect
HeH$^+$ in cool helium stars as those studied by Saio et al.~(2000).

Additionally, the discovery of the very metal-poor stars (Christlieb et
al.~2002, Frebel et al.~2005) strengthened the quest for
Population III stars.  It then follows that to study the evolution of
primordial molecules formed during post-recombination era is a
fundamental step to gain a better understanding of the ``how and when''
such stars were formed.

To establish the role of the main molecular species (LiH, LiH$^+$,
HeH$^+$, H$_2$, HD$^+$, HD) during the gravitational collapse which led
to the first stars formation, it is however well-known that their
opacity data are indeed needed. It is also just as important to know
the elementary mechanisms which lead to the formation and destruction
of such molecules and the corresponding rates that were present at the
conditions of the astrophysical environment. In other words, we
have to know as much as possible the chemical history of all the above
species.

An extensive analysis of the lithium chemistry has been reported in a
recent paper (Bovino et al. 2011) while an analogously accurate study
of the reactions which involve helium-based molecular species, and in
particular the HeH$^+$ molecular cation in reaction with H, is still
missing. The latter is indeed considered to be the first molecule
to appear in the Universe (Lepp et al. 2002):  in an environment
dominated by a low-density regime where 3-body collisions are unlikely
to occur, the radiative association is thus the main path to form that
molecule.  Hydrogen and helium, which are the most abundant species
which would have existed at the early universe epoch, can therefore
combine to form HeH$^+$ by the familiar reaction:
\begin{equation}
{\rm He} + {\rm H}^+ \rightarrow {\rm HeH}^+ + h\nu.
\end{equation}
for which the rate coefficient has been obtained from Zygelman et
al.~(1998) and fitted by Stancil et al.~(1998). On the other hand, the
main destruction path for the formed HeH$^+$ molecules is still the
reaction with H:
\begin{equation}
{\rm HeH}^+ + {\rm H} \rightarrow {\rm He} + {\rm H}_2^+,
\end{equation}
which however has never been studied with sufficient accuracy from a
theoretical point of view.  Crossed-beam experimental data  for
energies larger than 0.2 eV have been presented by Rutherford \&
Vroom~(1973), fitted by Linder et al.~(1995) and then included in the
early universe model calculations of Schleicher et al.~(2008), but no
direct calculations have been made available thus far.

In this paper we therefore report an accurate analysis, from ab initio
data, which deals with the above reaction.  We present new quantum
calculations for process (2) over a broad range of collision energies:
the corresponding rate coefficient has been obtained from the
calculated reactive cross sections and is presented in the following
sections, where newly computed abundance of HeH$^+$ is also
reported. Finally, we additionally compare it with the known behavior
of LiH$^+$ cations under the same environmental conditions and can
therefore draw some conclusions on their relative roles during the
early universe chemical evolution.

\section{The quantum reactive calculations}

The elementary reactions indicated in either direction by process (2)
have been studied both experimentally (Rutherford \& Vroom
1973, Karpas et al. 1979, Tang et al. 2005) and theoretically
(Aquilanti et al. 2000, Palmieri et al. 2000, Xu et al. 2008,
Ramachandran et al. 2009), because of its peculiar resonant features
detected experimentally.  Several potential energy surfaces (PES's)
have been obtained and employed, using quasi-classical trajectories,
close coupling (time-independent and time-dependent) methods to study
the inverse of the reaction (2).

The polar cation of the present study, HeH$^+$, has a large dipole
moment (1.66 Debye) and has about 11 vibrational levels
(Zygelman et al. 1998). For this reason it is therefore considered of
great importance in the early universe as possible coolant during the
post-recombination era.  As suggested by Dubrovich (1994), molecules
with large dipole moments could interact with the cosmic
microwave background radiation (CMB) inducing both spatial and spectral
distorsions (Maoli et al. 1994). The interaction with the CMB is
therefore crucial to determine the evolution and abundance of
primordial molecules. The study of the HeH$^+$ chemical evolution,
therefore, strongly depends on the accuracy of the main reaction rate
coefficients which lead to its destruction or formation. Formation
mechanisms for HeH$^+$ have already been extensively investigated
(Zygelman et al. 1998) while we wish here to focus on its destruction
mechanisms, and in particular on the direct exothermic (0.748 eV)
reaction (2).  In our calculations we employ one of the latest of the
available, calculated PES (Ramachandran et al. 2009). Since it is also
known that, under early universe conditions molecules are most likely
to be in their roto-vibrational ground state ($\nu$ = 0, $j$ = 0), we
have further decided to study reaction (2) starting from this level of
the target cationic molecule.

The exact quantum study of reactions involving an ionic system is
usually rather difficult due to the presence of a long range "tail" in
the interaction which therefore requires the use of a large basis set
to generate numerically converged cross sections and rates. The use of
an accurate computational method which is less costly in terms of
resources  is therefore mandatory.  In a recent paper (Tacconi et al.
2011) we have presented a novel computational implementation of the
negative imaginary potential (NIP) scheme proposed earlier by Baer and
coworkers (1990) and we shall use it for the present study.  We start
by solving the Schr\"odinger equation in a Body-Fixed (BF) frame of
reference, in the presence of an additional NIP. Hence we write the
Hamiltonian of the triatomic system in the BF frame, within the coupled
states (CS) (McGuire 1976) approximation

\begin{eqnarray}
\nonumber 
\hat{H} = -\frac{1}{2\mu}\nabla_R^2 +
\frac{\hat{J}^2+\hat{j}^2-2\hat{J}_z\hat{j}_z}{2\mu R^2} -
\frac{1}{2m}\nabla_r^2 + \frac{\hat{j}^2}{2mr^2} \\ +\hat{V}(r, R,
\vartheta) + \hat{V}^{\rm NIP}(r, R, \vartheta) 
\end{eqnarray} 
where $R$, $r$, and $\vartheta$ are the Jacobi coordinates, $\mu$ and
$m$ are the reduced masses of the triatomic complex and diatomic
molecule respectively.  $J$ and $\Omega$ are the total  angular
momentum and its projection on the relative axis while $j$ is the
diatomic rotational angular momentum. $\hat{V}(r, R, \vartheta)$ is the
interaction potential and $\hat{V}_{\rm NIP}$ has the functional form given
in Bovino et al. (2010b). The CS approximation provides the
exact dynamics when the target rotational angular momentum $j$ is
chosen to be equal to 0 (MgGuire 1976). Thus, it is here equivalent to
the full coupled-channel (CC) formulation. Because of the
flux-absorbing effect coming from the NIP, the resulting S-matrix is
now non-unitary and its default to unitarity yields here the
(state-to-all) reaction probability, as discussed in Tacconi et
al.~(2011).  From the reaction probability one can in turn obtain the
reactive cross section
\begin{equation} 
\sigma_{{\rm a}\rightarrow {\rm all}}(E) = \frac{\pi}{(2j_a + 1)
k_a^2}\sum_J\sum_{\Omega}(2J + 1) P^{J\Omega}_{{\rm a}\rightarrow {\rm all}}
\end{equation} 
while the rate coefficients are computed by averaging
the appropriate reactive cross sections over a Boltzmann distribution
of velocities for the incoming atom:  
\begin{equation} 
\alpha(T) = \frac{(8k_BT/\pi\mu)^{1/2}}{(k_BT)^2} 
\int_0^\infty\sigma_{{\rm a}\rightarrow {\rm all}}(E)\exp(-E/k_BT)E\,dE 
\end{equation} 
where $T$ is the gas temperature, $k_B$ is the Boltzmann constant and
$\mu$ the reduced mass of the system.  In the following section we
shall present the effects from the accurate rates computed above on the
final chemical evolution.

\section{The chemical network and the evolutionary modelling}

The evolution of the pregalactic gas is usually considered within the
framework of a Friedmann cosmological model and the primordial
abundances of the main gas components are taken from the standard big
bang nucleosynthesis results (Smith et al. 1993). The numerical values
of the cosmological parameters used in the calculation are obtained
from WMAP5 data (Komatsu et al.~2009) and are listed in Table~1 (for
details, see Coppola et al.~ 2011).

The abundance of HeH$^+$ is obtained by solving a set of differential 
coupled rate equations of the form
\begin{equation}
\frac{dn_i}{dt} = \alpha_{\rm form} -\alpha_{\rm dest}n_i,
\label{evol}
\end{equation}
where $\alpha_{\rm form}$ and $\alpha_{\rm dest}$ are formation
and destruction rates for the reactant species $i$ with number density
$n_i$, which in general depend on the number densities of the other
species and on the radiation field. The rate coefficients involved in
the helium chemistry network are the same as those reported in Galli
and Palla~(1998) (hereafter GP98), with the exception of the
destruction reaction (2) which indeed now comes from the present
quantum calculations. The additional equations which govern the
temperature and redshift evolutions have been reported in several
earlier papers (see e.g. GP98, Bovino et al.~2011) and therefore we
shall not discuss them within the present context.

\begin{table}
\begin{flushleft}
\caption{Cosmological model parameters. $H_0$ is the Hubble
constant, $\Omega_{\rm m}$, $\Omega_{\rm r}$, $\Omega_\Lambda$,
$\Omega_{\rm K}$ are density parameters, and $f_i$ are the initial
fractional abundances of the main species. $T_0$ is the present-day 
CMB temperature (see Coppola et al.~ 2011 for details).}
\begin{tabular}{ll}
\hline
Parameter & Numerical value\\
\hline 
$H_0$ & $70.5$~km~s$^{-1}$~Mpc$^{-1}$ \\
$T_0$ & $2.725$~K\\
$z_{\rm eq}$ & $3141$\\
$\Omega_{\rm dm}$ & $0.228$\\
$\Omega_{\rm b}$ & $0.0456$\\
$\Omega_{\rm m}$ & $\Omega_{\rm dm}+\Omega_{\rm b}$\\
$\Omega_{\rm r}$ & $\Omega_{m}/(1+z_{\rm eq})$\\
$\Omega_\Lambda$ & $0.726$\\
$\Omega_{\rm K}$ & $1-\Omega_{\rm r}-\Omega_{\rm m}-\Omega_\Lambda$ \\
$f_{\rm H}$ &  0.924\\
$f_{\rm He}$ & 0.076\\
$f_{\rm D}$ &  $2.38\times 10^{-5}$\\ 
$f_{\rm Li}$ &  $4.04\times 10^{-10}$\\ 
\hline
\end{tabular}
\vspace{1em}
\end{flushleft}
\end{table}

\section{Results and comparisons}

As already outlined in the Introduction, our understanding of the early
universe evolution strongly depends on our knowledge of the elementary
processes involving atomic and molecular species taken to be present at
that time.  At the molecular level, therefore the observables that one
needs to obtain are the various cross sections related to their
mechanisms of occurrence.  In the present work we have accurately
calculated the cross sections for HeH$^+$ ($^1\Sigma^+$) in reaction
with H($^1S$) using quantum dynamics and ab initio data which provide
the interaction forces.

In the following we shall report and discuss our results and
further compare the new HeH$^+$ abundances produced by the new quantum
reaction data  with those obtained for another molecular ion present in
the chemical network, LiH$^+$ (Bovino et al. 2011). 

\subsection{The reaction cross sections}

A set of three hundreds coupled equations which describe reaction (2)
have been solved using our recent, in-house code (Tacconi et al.
2011).  A direct product of a Colbert-Miller (Colbert \& Miller~1992)
discrete variable representation (DVR) of 100 points ranging from 
0.5~$a_0$ to 12~$a_0$ for the reagent ion vibrational coordinate and a
rotational basis of 40 spherical harmonics has been employed to
describe the reactive complex.  The solution matrix has been
propagated up to 50~$a_0$ and the convergence was checked as a function
of the propagator step-size, yielding a final convergence of cross
section values within about 1\%.

The calculations were carried out over a wide range of energies, from
10$^{-4}$ to 1.0 eV, and for total angular momentum values ($J$)
ranging  from 10 to 90 for the highest energy.  The NIP parameters were
tested following the Baer criteria (Baer et al. 1990): the NIP
stability has been reached for $r_{\rm min} = 2.75 a_0$, $r_{\rm max} =
5.75 a_0$ and the NIP order $n = 2$ (see Tacconi et al.~2011).

Our computed, $J$-converged integral cross sections are presented in
Figure 1. It is clear from the Figure the non-Langevin behavior of the
cross sections for this ionic reaction since, instead of becoming
independent of E at lower energies, they go through a marked maximum
around 10$^{-2}$ eV. In the same figure we report the experimental
(thick line, from Rutherford et al. 1973) and the extrapolated (dotted
line, from Linder et al. 1995) cross sections which reach down to
collision energy values around 10$^{-2}$ eV. The agreement between our
theoretical results and the experimental findings, both measured and
extrapolated, is reassuringly good and confirms the physical
reliability of the present NIP method for handling ionic reactions.  No
measured data are available at the lower collision energies as crossed
beam experiments with ionic partners become increasingly difficult in
that range. It is, however, worth noting here that the results of
Figure~1 are very different from the findings for LiH$^+$ (Bovino et
al. 2010a) in the same range of energies. Such differences are related
to different microscopic mechanisms of the reaction pathways between
the two systems, an aspect of the problem which we are planning to
analyse more in detail in further work. Qualitatively, the two systems,
although both undergoing no barrier, exothermic reactions, have very
different exothermicity values that affect product separations at low
energy.

\begin{figure}
\resizebox{\hsize}{!}{\includegraphics{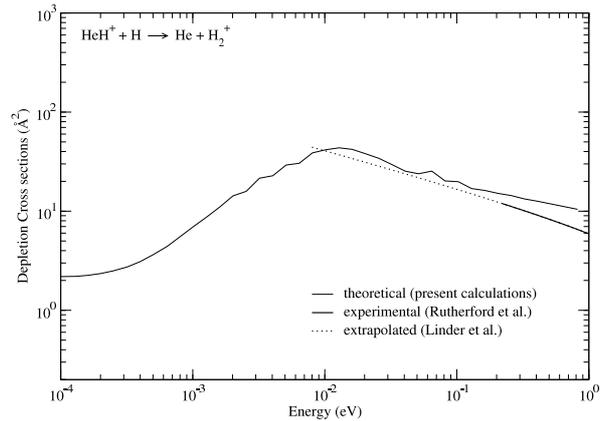}}
\caption{Computed integral cross sections as a function of
energy. The thick line represents the fitted experimental
data from Rutherford \& Vroom~(1973). The dotted line is
obtained from the formula given by Linder et al. (1995)
by extrapolating the Rutherford's data.}
\label{FigVibStab}
\end{figure} 

\begin{figure}
\resizebox{\hsize}{!}{\includegraphics{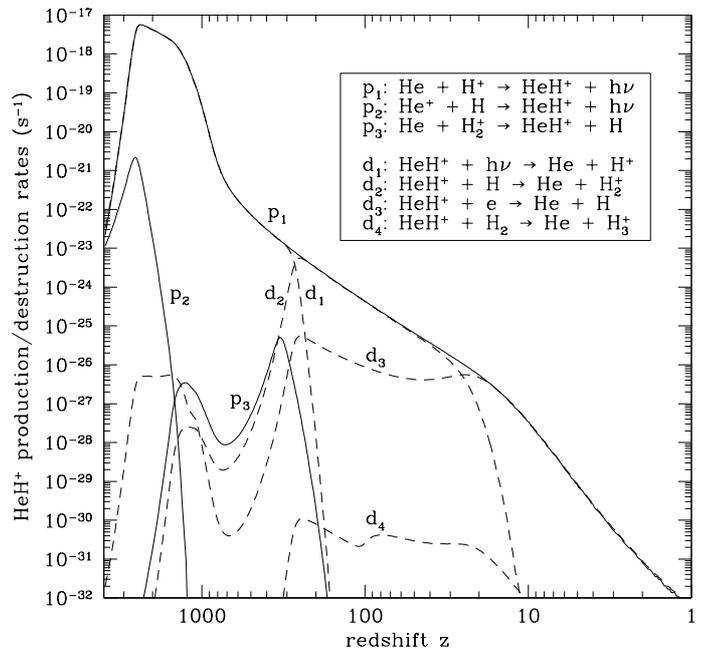}}
\caption{Computed production/destruction rates of HeH$^+$ as function 
of redshift. Solid and dashed curves represent production and 
destruction processes, respectively.}
\end{figure}

\subsection{The depletion rates}

The new rate coefficients are reported in Table 2 and further fitted
with the following formula:

\begin{equation}
	\alpha (T) = 4.3489\times10^{-10}T^{0.110373} e^{(-31.5396/T)}
\end{equation}
valid for $T \leq 1000$~K.  The experimental value obtained by Karpas
et al.~(1979) is also reported for comparison in the same Table. It is
again very reassuring to see that the only existing experimental datum
at 300 K is indeed very close to our computational value at the same
temperature.

In order to more clearly evaluate the significance of the newly
computed cross sections we need to view the process (2) within the
broader context of the variety of elementary events involved with this
molecular cation.  To that end, Figure~2 shows the main processes
leading to the formation and the destruction of HeH$^+$.  As clearly
shown by the figure, the radiative association process ($p_1$)
dominates the formation of HeH$^+$ over the whole range of redshifts,
while the association involving the He ion ($p_2$) and the one
involving the H$_2^+$ molecule ($p_3$), have negligible contributions
at the lower redshift values, the most important range for the present
study.

From the same figure one further realizes that the photodissociation
mechanism ($d_1$) is the dominant destruction path of the HeH$^+$ for
$z \geq 300$ while, on the other hand, the chemical path which involves
the reaction (2), the one which destroys HeH$^+$ ($d_2$) in collision
with H, is seen to be the most important process for $z \leq 300$.
A contribution also comes from the electron-assisted
dissociation reaction ($d_3$) that neutralizes the charged partners and
dominates over ($d_2$) for $z \leq 10$.

\begin{table}
\begin{flushleft}
\caption{Computed rate coefficient for the reaction (2) as a function of $T$}
\begin{tabular}{lll}
\hline
T(K) & $\alpha$~(cm$^3s^{-1})$ &\\
\hline 
1   & $4.41\times 10^{-12}$ & \\
5   & $2.11\times 10^{-11}$ & \\
10  & $5.47\times 10^{-11}$ & \\
20  & $1.33\times 10^{-10}$ & \\
30  & $2.11\times 10^{-10}$ & \\
50  & $3.43\times 10^{-10}$ & \\
70  & $4.39\times 10^{-10}$ & \\
80  & $4.76\times 10^{-10}$ & \\
100 & $5.33\times 10^{-10}$ & \\
200 & $6.76\times 10^{-10}$ & \\
300 & $7.39\times 10^{-10}$ & $9.10\pm 2.5\times 10^{-10}~(^\ast)$ \\
500 & $8.08\times 10^{-10}$ & \\ 
800 & $8.70\times 10^{-10}$ & \\ 
\hline
\end{tabular}
\vspace{1em}
\end{flushleft}
$(^\ast)$~experimental (Karpas et al.~1979)
\end{table}

\begin{figure}
\resizebox{\hsize}{!}{\includegraphics{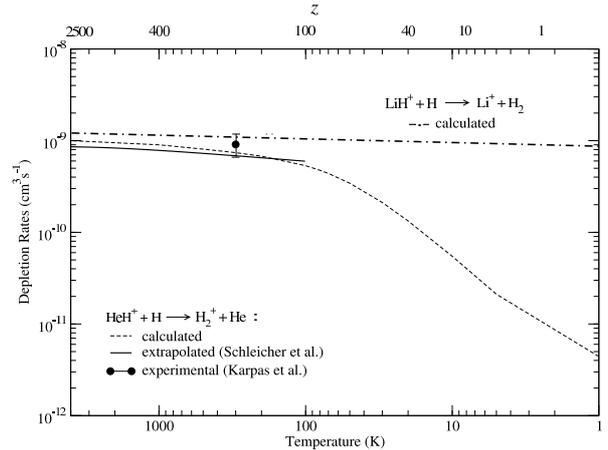}}
\caption{Depletion rate coefficients as a function of the temperature.
In the upper scale the redshift is also reported. The experimental rate
coefficients of Karpas et al.~(1979) and those extrapolated by
Schleicher et al.~(2008) are also presented. The lithium reactions rate
is from Bovino et al.~(2010a).}
\end{figure}

\subsection{A comparison with LiH$^+$}

We do know that during the Dark Ages of the universe, the existing
molecules interacted with CMB photons and with the present atoms.  What
we do not know is if these species existed with large enough abundances
to produce now detectable signals.  Because of its large dipole moment
(1.66 Debye), and the large abundances of its atomic components (H and
He), HeH$^+$, is certainly one of the prime candidates amenable to
observations (Persson et al. 2010).  This is even more important as our
recent ab-initio results (Bovino et al, 2009, 2010a, 2010b, 2011)
showed that lithium-containing molecules such as LiH and LiH$^+$,
in spite of their significant dipole moments that would favor
transitions (5.9 and 0.6 Debye), have very low abundances over the
whole range of redshifts studied.

In Figure 3 we report the calculated rate coefficients for the
depletion reaction (2) over a wide range of temperatures (redshifts)
together with the earlier calculations for LiH$^+$ depletion (Bovino et
al. 2010a).  As it is clear from that Figure, the HeH$^+$ + H reaction
turns out to be very efficient at the redsfhifts above 100, with
rate values comparable to those of the LiH$^+$ + H destruction reaction
given in the same figure (dot-dashed line).  However, while the latter
remains essentially constant over the whole range of examined
redshifts, the efficiency of reaction (2) quickly decreases for $z \leq
100$,  coming down by about two orders of magnitude by the time unitary
redshift values are reached. This behavior therefore favors the HeH$^+$
as a more likely survival candidate vis-\'a-vis the LiH$^+$ case,
thereby providing a good molecular coolant acting during the primordial
star formation.  In the same figure we also report the experimental
rate at 300 K given by Karpas et al.~(1979); the rates included
in the GP98 model are taken from the experimental data of Rutherford \&
Vroom (1973). As seen earlier for the cross sections, the experimental
rates are in good agreement with our quantum results (see also Table
2).

In Figure 4 we report the calculated abundances for HeH$^+$ obtained
using our new quantum rates discussed above: they are compared with the
old ones obtained by GP98 and also with our newest results for LiH$^+$
(Bovino et al., 2011).  The scenario emerging from the above
calculations indicates that for redshift around $z\simeq 20$, the
molecular HeH$^+$ abundance has now increased by more than one order of
magnitude with respect to the earlier calculations. It thus remains a
more abundant ionic species LiH$^+$, hence becoming an
interesting molecular candidate for actual detection. As shown by
Maoli et al.~(1994) (see also Schleicher et al.~2008), the relevant
quantity for determining the imprint of molecules on the CMB is the
optical depth due to line absorption. In the case of HeH$^+$, the
optical depth has a broad peak around a frequency of $\sim 50$~GHz, due
to redshifted rotational transitions with $j=4$-6 of the ground
vibrational level. With the new rate coefficient derived in this work,
the maximum value of the optical depth, $\sim 10^{-7}$, is about one
order of magnitude larger than the value computed earlier by Schleicher
et al.~(2008), an encouraging result for observational perspectives.
The new reaction rate has also an effect on the abundance of H$_2^+$,
that is reduced by a factor of $\sim 5$ in the range $10 \lesssim z
\lesssim 50$ with respect to earlier results. However this reduction
has no significant effect on the abundance of H$_2$ because the H$_2^+$
channel of production of H$_2$ is no longer active at these
redshifts.

\begin{figure}
\resizebox{\hsize}{!}{\includegraphics{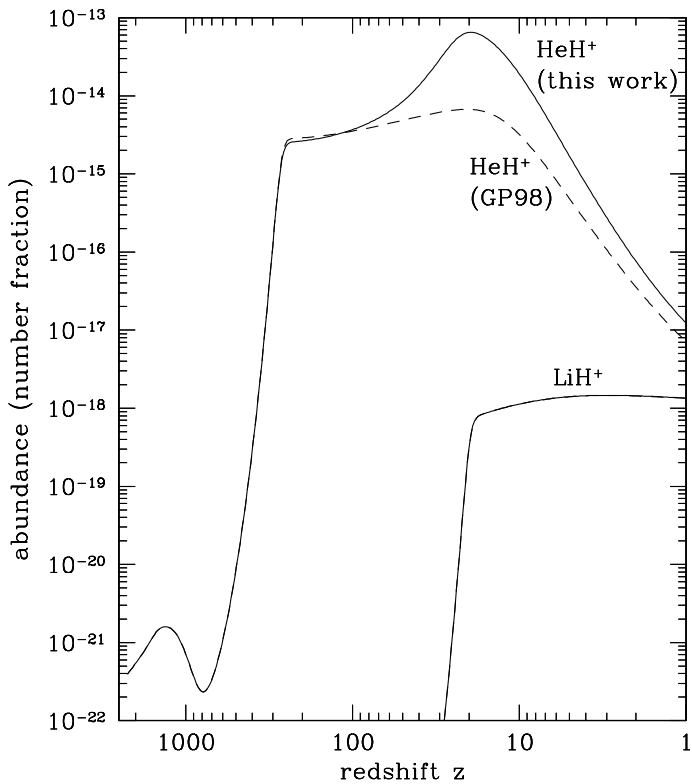}}
\caption{Computed relative abundances of HeH$^+$ in the
post-recombination era as function of redshift $z$: present results
(solid curves), results of GP98 (dashed curves). The LiH$^+$ results
are from Bovino et al. (2011).}
\end{figure}

\section{Conclusions}

We have carried out a new set of ab initio, quantum calculations for
the rate coefficient of the reaction (2), which therefore permit us a
revision of the abundance of HeH$^+$ molecule within the expected
conditions operating during the early universe.  One of the main
results from the present reactive calculations is that the chemical
destruction rate of this polar cation exhibits a temperature dependence
that turns out to be very different from previous data, both
experimental and theoretical, especially for temperatures smaller than
100 K.

These new results obtained for the HeH$^+$ abundance now indicate that
this molecule should be much more likely to have survived at low
redshift, since its fractional abundance goes up by more than one order
of magnitude from previous estimates, becoming of the order of
10$^{-13}$. This new findings should be considered within the broader
context whereby the fractional abundance of LiH$^+$, recently
calculated by Bovino et al. (2011), turns out to be significant mostly
in the low-redshift region of $z \lesssim 30$.  The comparison between
the specific fractional abundances of the two species therefore
suggests rather compellingly that the HeH$^+$ ionic molecule, if more
abundant than the LiH$^+$, would become more likely to be observed
experimentally, even more so than what was suggested by previous works
(GP98, Persson et al. 2010).

It is also worth to mention that, due to its larger dipole moment
and to its greater number of roto-vibrational levels, when compared
with LiH$^+$, the molecular cation HeH$^+$ turns out to be the best
candidate to date for experimental observations in different
astrophysical environments, a result now supported by the present
ensemble of accurate computational studies from quantum, ab initio
methods.

\acknowledgements

We thank the CINECA and CASPUR consortia for providing us with the
necessary computational facilities and the University of Roma
``Sapienza'' for partial financial support. We are grateful to Dr. D.
De Fazio for providing us with the potential energy surface of
Ramachandran et al. (2009). D.G. also thanks F. Palla for discussions
on this subject.

\bigskip
\bigskip
\noindent {\large\bf References}
\bigskip

\noindent Aquilanti V. et al., 2000, Chem. Phys. Lett., 318, 619 \\
\noindent Baer M., Ng C. Y., Neuhauser D., \& Oreg Y., 1990, J. Chem. Soc. Faraday Trans., 86, 1721 \\
\noindent Bovino S., Wernli M. \& Gianturco F. A., 2009, ApJ, 699, 383 \\
\noindent Bovino S., Stoecklin T. \& Gianturco F. A., 2010a, ApJ, 708, 1560 \\
\noindent Bovino S., Tacconi M., Gianturco F. A. \& Stoecklin T., 2010b, ApJ, 724, 106 \\
\noindent Bovino S., Tacconi M., Gianturco F. A., Galli D., Palla F., 2011, \apj, in press \\
\noindent Christlieb N. et al., 2002, Nature, 419, 904 \\
\noindent Colbert D. T., and Miller W. H., 1992, J. Chem. Phys., 96, 1982 \\
\noindent Coppola, C.~M., Longo, S., Capitelli, M., Palla, F., \& Galli, D.\ 2011, \apjs, 193, 7  \\
\noindent Dubrovich V. K., 1993, Astron. Lett., 19, 53 \\
\noindent Engel E. A., Doss  N., Harris G. J., and Tennyson J., 2005, \mnras, 357, 471 \\
\noindent Frebel A. et al., 2005, Nature, 434, 871 \\
\noindent Galli D., Palla F., 1998, A\&A, 335, 403 (GP98) \\
\noindent Harris G. J., Lynas-Gray A. E., Miller S., and Tennyson J., 2004, \apj, 617, L143 \\
\noindent Karpas Z., Anicich V., and Huntress, Jr. W. T., 1979, J. Chem. Phys., 70, 2877 \\
\noindent Komatsu, E., et al. 2009, ApJS, 180, 330 \\
\noindent Lepp S., Stancil P. C. \& Dalgarno A., 2002, J. Phys. B: At. Mol. Opt. Phys., 35, R57-R80 \\
\noindent Linder F., Janev R. K., and Botero J., 1995, Atomic and Molecular Porcesses in Fusion Edge Plasmas, ed. R. K. Janev (NY: Plenum Press), 397 \\
\noindent Liu X.-W. et al., 1997, \mnras, 290, L71 \\
\noindent Maoli R. et al., 1994, \apj 425, 372 \\
\noindent McGuire P., 1976, Chem. Phys., 13, 81 \\
\noindent Palmieri P. et al., 2000, Mol. Phys., 98, 1835 \\
\noindent Persson, C.~M., et al.\ 2010, \aap, 515, A72 \\
\noindent Ramachandran C. N., De Fazio D., Cavalli S., Tarantelli F., and Aquilanti V., 2009, Chem. Phys. Lett., 469, 26  \\
\noindent Roberge W., and Dalgarno A., 1982, ApJ, 255, 489 \\
\noindent Rutherford J. A., and Vroom D. A., 1973, J. Chem. Phys., 58, 4076 \\
\noindent Saio H., and Jeffery C. S., 2000, \mnras, 313, 671 \\
\noindent Schleicher, D.~R.~G., Galli, D., Palla, F., Camenzind, M., Klessen, R.~S., Bartelmann, M., \& Glover, S.~C.~O.\ 2008, \aap, 490, 521 \\
\noindent Smith M. S., Kawano L. H. \& Malaney R. A., 1993, ApJS, 85, 219 \\
\noindent Stancil P. C., Lepp S., and Dalgarno A., 1998, \apj, 509, 1 \\
\noindent Tacconi M., Bovino S., and Gianturco F. A., Rendiconti Lincei, in press \\
\noindent Tang X. N. et al., 2005, J. Chem. Phys., 122, 164301 \\
\noindent Zygelman B., Stancil P. C., and Dalgarno A., 1998, \apj, 508, 151 \\
\noindent Xu W., Liu X., Luan S., Zhang Q., and Zhang P., 2008, Chem. Phys., Lett., 464, 92   \\

\end{document}